\documentclass[12pt]{article}

\usepackage[intlimits]{amsmath}
\usepackage[cp1251]{inputenc}
\usepackage[T2A]{fontenc}
\usepackage[russian,english]{babel}

\usepackage[dviwin]{graphicx}
\usepackage{amsfonts}

\begin{document}
\title{Dimension two vacuum condensates \\in gauge-invariant theories}
\author{D.V.Bykov$^{\ddag}$, A.A.Slavnov$^{\ddag,\natural}$}
\date{02.04.2005}
\maketitle

\language 0

$\ddag$ Moscow State University, Physics Faculty, Leninskie gory,
bld 1-2, GSP-2, 119992 Moscow, Russia

$\natural$ V.A.Steklov Mathematical Institute, Gubkina str., bld.
8, GSP-1, 119991 Moscow, Russia

\begin{center}
\textbf{Abstract}
\end{center}

Gauge dependence of the dimension two condensate in Abelian and
non-Abelian Yang-Mills theory is investigated.

\section{Introduction}

Recently much attention has been drawn to vacuum condensates
$<0|A_{\mu}^{a 2}|0>$ and $<0|\overline{c}^{a}c^{a}|0>$ in
non-Abelian gauge theories. It is believed that these condensates
carry information about nonperturbative phenomena in quantum
chromodynamics, such as quark confinement
~\cite{gubarev,stodolsky}. They contribute to the nonperturbative
parts of the gluon ~\cite{lavelle,boucaud} and the quark
~\cite{arriola} propagators. In the papers
~\cite{gubarev,stodolsky} it was suggested that the gluon
condensate may be sensitive to various topological defects such as
Dirac strings and monopoles. Considered condensates are vacuum
expectation values of gauge dependent operators, which makes
problems for calculation of observable effects. In the papers
~\cite{slavnov,slavnov2} it was shown, that if one considers
Yang-Mills theory as a limit of a (regularized) noncommutative
gauge-invariant theory, then the v.e.v. $<\int d^{4}x
A_{\mu}^{2}>$ doesn't depend on the choice of gauge and,
therefore, it may have a direct physical meaning. This proof
essentially depends on the existence of a gauge-invariant
regularization of noncommutative theories, and this question
requires further investigation. Thus, it is interesting to explore
the gauge invariance of dimension 2 condensate in the commutative
theory, and to study the question of its possible contribution to
the Wilson OPE. A partial answer to this question for the case of
Abelian theory was given in the work ~\cite{slavnov}. In this work
we continue investigating this question in the Abelian, as well as
in non-Abelian cases, and we explore the question of Wilson OPE in
noncommutative theory.

\section{Some condensates of mass dimension two and their applications in field theory}

In this section we will be dealing with Green's functions and
v.e.v.'s of the gluon field $A_{\mu}(x)$ in $\alpha$-gauges, and
of the ghost fields $\overline{c}(x)$, $c(x)$.

The simplest Green's functions in non-Abelian gauge theories are
the functions \\ $<T A_{\mu}(x) A_{\nu}(y)>$, $<T \overline{c}(x)
c(y)>$. Numerical calculations of path integrals, determining
these v.e.v.'s, allow to explore the nonperturbative contributions
to these propagators. These contributions play an important role
in the Wilson expansions of operators $T A_{\mu}(x) A_{\nu}(y)$
and $T \overline{c}(x) c(y)$, where they appear as the power
corrections to the leading term of order $O((x-y)^{-2})$, which
corresponds to the unit operator in the expansion. The next
operators, which contribute to this expansion are the operators of
mass dimension 2: $A_{\mu}(x) A^{\mu}(x)$ and $\overline{c}(x)
c(x)$. In this work we will be dealing with these condensates.
Their contribution to the Wilson OPE has the following form
~\cite{kondo}:
\begin{eqnarray}
\int d^{4}x e^{ipx} (T A_{\mu}^{a}(x) A_{\nu}^{b}(0)) \underset{p
\rightarrow \infty}{\rightarrow} C^{[1] ab}_{\mu\nu}(p) \cdot 1 +
C^{[A_{\rho}^{2}]ab}_{\mu\nu}(p) (A_{\rho}^{c})^{2} +
C^{[\overline{c} c] ab}_{\mu\nu}(p) \overline{c}^{d} c^{d} + ...
\end{eqnarray}
If one is interested in the behavior of the gluon propagator at
large momenta, then one should take the v.e.v. of this expression,
which obviously depends on the values of the condensates
$<0|(A_{\rho}^{c})^{2}|0>$ and $<0|\overline{c}^{d} c^{d}|0>$. In
particular, it is of interest to know if it is possible to
construct gauge-invariant or at least BRST-invariant combinations
of these condensates (or corresponding operators). The answer to
this question was partially given in the work ~\cite{kondo}: in
gauges with the gauge fixing/ghost term of the form\footnote{Here
$B \;$ is an auxiliary Nakanishi-Lautrup field, and integration
over it can be easily done in the path integral}
\begin{eqnarray}\label{300}
L_{GF+FP} = \frac{\alpha'}{2} B^{a} B^{a} - \frac{\alpha'}{2} g
t^{abd} c^{b}\overline{c}^{d} B^{a} + B^{a}
\partial_{\mu}A_{\mu}^{a} + \overline{c}^{a}M_{ab}c^{b} -
\frac{\alpha'}{8} g^{2} t^{abd}t^{aef}
\overline{c}^{b}\overline{c}^{d}c^{e}c^{f}
\end{eqnarray}
there exists a BRST-invariant operator $\mathfrak{O} \equiv \int
d^{4}x (\frac{1}{2} A_{\mu}(x)A^{\mu}(x) - \alpha' \overline{c}(x)
c(x) )$. It is worth noting that BRST-invariance of this operator
is preserved in the U(1)-theory in Lorentz-type gauges, but in the
general case of Yang-Mills theory in the widely used Lorentz-type
gauge (the so-called $\alpha$-gauge) this statement is no longer
correct. No value of $\alpha'$ can turn $L_{GF+FP}$ into
$$L_{GF+FP} = - \frac{\alpha}{2} B^{a} B^{a} + B^{a}
\partial_{\mu}A_{\mu}^{a} + \overline{c}^{a}M_{ab}c^{b},$$ which corresponds to Lorentz-type gauge fixing. Besides,
one can directly check, that
\begin{eqnarray}
\delta \mathfrak O = \int d^{4}x \left[
\partial_{\mu}(A_{\mu}^{a}c_{a}) + \frac{\alpha}{2} t^{abd} c^{b} c^{d}
\overline{c}^{a}\right] = \int d^{4}x \frac{\alpha}{2} t^{abd}
c^{b} c^{d} \overline{c}^{a} \neq 0.
\end{eqnarray}
On the other hand, in the physical sector the operator $\mathcal A
\equiv c^{b} c^{d} \overline{c}^{a}$ is equivalent to the
null-operator because of the non-zero ghost
number:$$|\psi_{phys}>: \; Q_{ghost}|\psi_{phys}> =
Q_{BRST}|\psi_{phys}>=0$$
\begin{equation}\label{80}
0 = <\psi_{phys}^{1}|[iQ_{ghost},\mathcal A]|\psi_{phys}^{2}> =
<\psi_{phys}^{1}|\mathcal A|\psi_{phys}^{2}>
\end{equation}
Note that due to BRST invariance of the above mentioned operator
in gauges, determined by the functional (\ref{300}), and also in
Abelian theory, it is easy to see, that BRST-invariance is
transferred to the case of maximal Abelian gauge, which has an
Abelian sector analogous to the U(1)-theory (more precisely, the
$U(1)^{N-1}$-subgroup of the SU(N) group), and in the non-Abelian
sector the gauge is fixed by a functional of type (\ref{300}). It
has been also proved in \cite{kondo}.

Let's mention one important difference between the propagators of
gauge fields in the Abelian and non-Abelian theories. In Abelian
U(1)-theory the gauge parameter $\alpha$ enters the full
propagator of the photon field only through the trivial
longitudinal part:
\begin{equation}\label{90}
G_{\mu\nu}(p) = (\eta_{\mu\nu}-\frac{p_{\mu}p_{\nu}}{p^{2}})
G(p^{2}) - \alpha \frac{p_{\mu}p_{\nu}}{p^{4}},
\end{equation}
where $G(p^{2})$ doesn't depend on $\alpha$ (in non-Abelian theory
this statement is false, which is easy to show, making
calculations in lowest orders of perturbation theory). The proof
of this statement can be found, for instance, in ~\cite{slavnov}.
In the present article a proof, based on the Ward identities, is
given in Appendix 1. This observation allows us to build the
following gauge-invariant quantity:
\begin{equation}\label{91}
<0|T(A_{\mu}(x)A^{\mu}(y)+ \alpha \overline{c}(x)c(y))|0>
\end{equation}
In Abelian theory the ghost field is free, and $$<0|T
(\overline{c}(x)c(y))|0> = \int d^{4}p \; e^{ip(x-y)}
\frac{1}{p^{2}+i\epsilon}.$$ Taking into account that
$$<0|T(A_{\mu}(x)A^{\mu}(y))|0> = \int d^{4}p \; e^{ip(x-y)}
G_{\mu}^{\mu}(p),$$ we prove gauge invariance of the quantity
(\ref{91}). From analogous considerations it is clear that the
vacuum condensate $<0|A_{\mu}(x)A^{\mu}(x)+ \alpha
\overline{c}(x)c(x)|0>$ is gauge-invariant, too.

\section{Contribution of condensates $<A_{\mu}^{2}>$ and $<\overline{c} c>$\\ to the Wilson expansion of the operator\\ $\mathcal K(x,y)
\equiv T(A_{\mu}(x)A^{\mu}(y) + \alpha \overline{c}(x) c(y))$ in
U(1)-theory}

In this section we discuss the question, how gauge invariance of
the vacuum expectation value $<0|\mathcal K(x,y)|0>$ constrains
the possible terms, contributing to the Wilson expansion of this
operator. The Wilson expansion of the operator $\mathcal K(x,y)$
in commutative $U(1)$-theory (including interaction with a spinor
field $\psi(x)$ of mass $m$) takes the following form\footnote{We
don't take into account possible condensates with non-zero ghost
number, because they don't contribute to the physical sector of
the theory.}:
\begin{eqnarray}\label{899}
\mathcal K(x,y) \underset{x-y\rightarrow 0}{\rightarrow}
C_{0}(x-y) \cdot 1 + C_{1\mu\nu}^{(1)}(x-y) A^{\mu}(y)A^{\nu}(y)
+\\ \nonumber + C_{2 ab}^{(1)}(x-y) \overline{c}^{a}(y)c^{b}(y)+
...
\end{eqnarray}
Coefficients $C^{(1)}_{1 \mu\nu}$ and $C_{2 ab}^{(1)}$ are
dimensionless. Besides, we have just two possible tensor
structures: $\eta_{\mu\nu}$ and $z_{\mu} \equiv x_{\mu} -
y_{\mu}$. There's one obvious mass scale in the theory - the
spinor mass $m$, and also a hidden one - the subtraction point
$\mu$. However, the latter enters only logarithmic corrections to
the power expansion (we're considering the case $z \rightarrow 0$,
but for this expansion to be valid it is necessary that
$|g\ln{z^{2}\mu^{2}}|<1$). Thus, any coefficient function $C(z)$
of the expansion has the following structure:
\begin{equation}\label{45}
C(z) = A z^{\alpha} (1 + \sum_{k=1}^{\infty} A_{k} (g
\ln{z^{2}\mu^{2}})^{k})
\end{equation}
The question of spinor mass is a bit more subtle. Can the terms
with negative powers of $m$ contribute to the coefficient
functions? If yes, then the whole structure of the Wilson
expansion for nonsingular terms is destroyed, because then it
would be possible to take operators of arbitrary dimension and to
obtain operators of other mass dimension by simply dividing them
with the necessary power of mass. As a result in any order of $z$
we would get an infinite series of condensates. Fortunately, here
we can apply Weinberg's theorem, which guarantees the existence of
a limit of zero spinor mass for quantum electrodynamics in case of
diagrams without explicit external momenta, and, therefore, the
terms mentioned above are prohibited.

Taking into account all what was said above, let's write down the
most general form of the coefficient functions $C^{(1)}_{1
\mu\nu}$ и $C_{2 ab}^{(1)}$:
\begin{equation}\label{500}
C^{(1)}_{1 \mu\nu} = \beta \eta_{\mu\nu}; \; C_{2 ab}^{(1)} =
\kappa \delta_{ab};
\end{equation}

Substituting this in the expansion (\ref{899}), we obtain:
\begin{equation}\label{999}
\mathcal K(x,y) \underset{x-y\rightarrow 0}{\rightarrow}
C_{0}(x-y) \cdot 1 + \beta A^{2}(y) +\kappa
\overline{c}^{a}(y)c^{a}(y)+ ...
\end{equation}

Gauge invariance of the v.e.v. of the l.h.s. of this equation
constrains the coefficients $\beta$ and $\kappa$:
\begin{equation}\label{777}
\frac{\kappa}{\beta} = \alpha
\end{equation}

This equality doesn't require any special proof, as we showed in
the previous section that the vacuum condensate
$<0|A_{\mu}^{2}(x)+\alpha \overline{c}(x)c(x)|0>$ is gauge
independent in the Abelian case.

\section{A necessary consequence of $\frac{d}{d \alpha} <A_{\mu}^{2}>=0$}

The non-Abelian case is especially interesting, though. In the
work ~\cite{slavnov} with the help of noncommutative field theory
methods it was shown, that the condensate
$<0|A_{\mu}^{a}(x)A^{\mu}_{a}(x)|0>$ is gauge-invariant, but this
statement requires some extra explanation, because, for example,
from (\ref{90}) it is clear, that in the particular case of
Abelian theory
$$\frac{d}{d\alpha}<0|A_{\mu}(x)A^{\mu}(x)|0> = D(0),$$ where $$D(x) = - \int \frac{e^{ipx}}{p^{2}+i\epsilon} d^{4}p,$$ and this
quantity is equal to zero only if $D(0)=0$, which is true, for
instance, in dimensional regularization. It is quite probable,
that in noncommutative theory the requirement of the existence of
a gauge-invariant regularization is rather constraining, and the
condition mentioned above is a necessary condition for it. This
question is non-trivial and requires further investigation. Here
we present one of the possible methods for checking the equality
\begin{equation}\label{92}
\frac{d}{d \alpha} <A_{\mu}^{2}> = 0,
\end{equation}
and more precisely we will show that there's a necessary
condition:
\begin{equation}\label{93}
<\overline{c}^{a}(x)c^{a}(x)>|_{\alpha=0}=0,
\end{equation}
i.e. the absence of ghost condensation in the Lorentz gauge.

The path integral, which determines the vacuum condensate, has the
following form in Yang-Mills theory:
\begin{eqnarray}
<\int d^{4}x A_{\mu}(x)A^{\mu}(x)> = {\rm N^{-1}} \int \left(\int
d^{4}x
A_{\mu}(x)A^{\mu}(x)\right) \exp{\{i\int d^{4}x ({\rm L}(A_{\mu}) +}\\
\nonumber { + \frac{1}{2\alpha} (\partial_{\mu} A^{\mu})^{2} +
\overline{c}^{a} M_{ab} c^{b}) \}} (\underset{x,
\mu}{\prod}dA_{\mu}(x) d\overline{c} dc),
\end{eqnarray}
where $\rm L$ is the Lagrangian for Yang-Mills fields without
matter fields:
\begin{eqnarray}
  {\rm L }&=& -\frac{1}{4} {\rm tr }(F^{\mu\nu}F_{\mu\nu}) \\
  F_{\mu\nu} &=& \partial_{\nu} A_{\mu} - \partial_{\mu} A_{\nu} +
g [A_{\mu},A_{\nu}]
\end{eqnarray}
Let's make the transformation
$$A_{\mu} \rightarrow A_{\mu} - \frac{\delta \alpha}{2 \alpha} D_{\mu}^{x} \int M^{-1}(x,y) \partial^{\nu}A_{\nu}(y) d^{4}y $$ in this path integral, and, keeping only the terms of order $\delta \alpha$,
we obtain:
\begin{eqnarray} \label{15}
<\int\limits_{\alpha} d^{4}x A_{\mu}(x)A^{\mu}(x)> =
<\int\limits_{\alpha + \delta \alpha} d^{4}x A_{\mu}(x)A^{\mu}(x)>
- \\ \nonumber - \frac{\delta\alpha}{\alpha} \int d^{4}x \int e^{i
S} {\rm tr}(A_{\mu}(x) D_{\mu}^{x} (M^{-1}
\partial_{\nu} A^{\nu})) {\prod}dA_{\mu}(x) d\overline{c} dc.
\end{eqnarray}
Let's take into account that
\begin{eqnarray}
{\rm tr}(A_{\mu}(x) D_{\mu}^{x} (M^{-1} \partial_{\nu}
A^{\nu})^{x}) = A_{\mu}^{a}(x) D_{\mu}^{ab} (M^{-1}_{bc}
\partial_{\nu} A^{\nu}_{c})^{x} = \\ \nonumber =A_{\mu}^{a}(x) (\delta^{ab}
\partial_{\mu} - t^{acb} A_{\mu}^{c}(x)) (M^{-1}_{bc}
\partial_{\nu} A^{\nu}_{c})^{x}
\end{eqnarray}
and the antisymmetry of the structure constants $t^{abc}$ of the
gauge group:
\begin{equation}\label{14}
A_{\mu}^{a}(x) A_{\mu}^{c}(x) t^{abc} = \frac{1}{2}(A_{\mu}^{a}(x)
A_{\mu}^{c}(x) t^{abc} + A_{\mu}^{c}(x) A_{\mu}^{a}(x) t^{cba}) =
0.
\end{equation}
Therefore (\ref{15}) takes the following form after integration by
parts in the last term:
\begin{eqnarray} \label{16}
<\int\limits_{\alpha} d^{4}x A_{\mu}(x)A^{\mu}(x)> =
<\int\limits_{\alpha +  \delta \alpha} d^{4}x
A_{\mu}(x)A^{\mu}(x)> + \\ \nonumber + \frac{\delta\alpha}{\alpha}
\int d^{4}x d^{4}y \int e^{i S}
\partial^{\mu}A_{\mu}^{a}(x) M^{-1}_{ab}(x,y)
\partial^{\nu} A_{\nu}^{b}(y) {\prod}dA_{\mu}(x) d\overline{c} dc,
\end{eqnarray}
where $M(x,y)$ is the kernel of the operator $M$. Thus, we get:
\begin{equation}\label{400}
\frac{d}{d \alpha} <\int d^{4}x A_{\mu}^{2}(x)> =
-\frac{1}{\alpha} \int d^{4}x d^{4}y \int
\partial_{\mu}A_{\mu}^{a}(x) M^{-1}_{ab}(x,y)
\partial_{\nu}A_{\nu}^{b}(y) e^{iS} \prod dA
d\overline{c} dc
\end{equation}

The r.h.s. of the equation (\ref{400}) can be rewritten in the
following form:
\begin{eqnarray}\label{401}
\frac{d}{d \alpha} <\int d^{4}x A_{\mu}^{2}(x)> =
-\frac{1}{\alpha} \int d^{4}x \;d^{4}y
<\partial_{\mu}A_{\mu}^{(0),a}(x)
\partial_{\nu} A_{\nu}^{(0),b}(y)> <M^{-1}_{ab}(x,y)>  - \\
\nonumber - \frac{1}{\alpha} \int d^{4}x\; d^{4}x'\; d^{4}y \;
d^{4}y'
\partial_{\mu} D_{\mu\mu'}^{(0),ac}(x-x') \;
\partial_{\nu}D_{\nu\nu'}^{(0),bd}(y-y')T_{\mu'\nu'}^{abcd} (x',y').
\end{eqnarray}
Here $A_{\mu}^{(0),a}$ is the free gauge field, аnd
$D_{\mu\mu'}^{(0),ab}(x-y)$ is its propagator. The function
$T_{\mu'\nu'}^{abcd}$ can be written as a perturbation theory
series, using formula (\ref{400}). It is only important, that this
function is nonsingular at $\alpha=0$. At the same time the free
Green's function satisfies the equation
\begin{equation}\label{402}
\partial_{\mu}^{x} D_{\mu\nu}^{(0),ab}(x-y) = -\alpha  \delta_{ab} \partial_{\nu}^{x} \int
\frac{e^{ik(x-y)}}{k^{2}+i\epsilon} d^{4}k
\end{equation}
It follows that in the limit $\alpha \rightarrow 0$ the second
term in the r.h.s. of (\ref{401}) is equal to zero, and
\begin{equation}\label{403}
\frac{d}{d \alpha} <\int d^{4}x \; A_{\mu}^{2}(x)>|_{\alpha=0} = <
\int d^{4}x \; M_{aa}^{-1} (x,x, A)>|_{\alpha=0} =
<\overline{c}^{a}(x)c^{a}(x)>|_{\alpha=0}
\end{equation}

In this way we obtain the necessary condition for the gauge
independence of the condensate $<\int d^{4}x
A_{\mu}(x)A^{\mu}(x)>$:
\begin{equation}\label{180}
<\int dx \; \overline{c}^{a}(x) c^{a}(x)>|_{\alpha=0} = 0
\end{equation}
In the Abelian case this requirement reduces to the condition
$D(0)=0$, which was discussed above. Let us explain once again,
that the equation (\ref{180}) should hold in commutative theory,
if it is obtained as a limit $\xi \rightarrow 0$ from a
gauge-invariant regularized noncommutative theory. It is worth
noting, that gauge invariance of the condensate $<A_{\mu}^{2}>$ in
non-Abelian theory doesn't guarantee its appearance in the Wilson
expansion of a gauge-invariant product of operators. The fact is
that gauge invariance of this condensate was proved by using the
noncommutative formulation of the theory at an intermediate stage.
At the same time, as it is shown in Appendix 2, in noncommutative
theory the Wilson expansion may be violated by terms of order
$\xi$, which doesn't allow us to make the conclusion about the
appearance of this operator in the Wilson expansion.

\section{Conclusion.}

In this work we have obtained the following results: first of all,
in $U(1)$-theory we showed the gauge invariance of the v.e.v.
$<0|T(A_{\mu}(x)A^{\mu}(y) + \alpha \overline{c}(x) c(y))|0>$ and
the explicit form of the Wilson expansion of this operator was
built. We suggested a method for checking the equation
$\frac{d}{d\alpha} <A_{\mu}^{a 2}> = 0$ in the non-Abelian theory
by studying the properties of the ghost condensate. It is shown,
that in noncommutative field theory the Wilson OPE is no longer
valid, and therefore gauge invariance of the condensate
$<A_{\mu}^{2}>$ doesn't guarantee, that it contributes to the
Wilson expansion of the product of gauge-invariant operators in
commutative theory.\newline

\textbf{Appendix 1.}

\emph{Proof of the statement} $G_{\mu\nu}(p) =
(\eta_{\mu\nu}-\frac{p_{\mu}p_{\nu}}{p^{2}}) G(p^{2}) - \alpha
\frac{p_{\mu}p_{\nu}}{p^{4}}; \; \frac{d G(p^{2})}{d\alpha}=0$
\begin{equation}\label{210}
\frac{d}{d \alpha} <T(A_{\mu}(x)A^{\mu}(0))> = \frac{d}{d\alpha}
\left[ \frac{\int A_{\mu}(x) A^{\mu}(0)  e^{i S(\alpha)}
\underset{x}{\prod}dA d \overline{c} dc}{\int e^{iS(\alpha)}
\underset{x}{\prod} dA d\overline{c} dc} \right]
\end{equation}
It is convenient to introduce the following notation:
\begin{eqnarray}
\nonumber {\rm Q}(\alpha) \equiv \int A_{\mu}(x) A^{\mu}(0)  e^{i
S(\alpha)}
\underset{x}{\prod}dA d \overline{c} dc;\\
\nonumber {\rm N}(\alpha) \equiv \int e^{iS(\alpha)}
\underset{x}{\prod} dA d\overline{c} dc; \\ {\rm C}(x) \equiv
<T(A_{\mu}(x)A^{\mu}(0))>;
\end{eqnarray}
Then
\begin{equation}\label{220}
\frac{d {\rm C}}{d\alpha} = {\rm \frac{ Q'\;N-Q\;N'}{N^{2}}}
\end{equation}
\begin{equation}\label{230}
{\rm Q'} = \int A_{\mu}(x)A^{\mu}(0) \left(-\frac{i}{2 \alpha^{2}}
\int (\partial_{\mu}A^{\mu})^{2} d^{4}y\right) e^{iS} \prod dA
d\overline{c} dc
\end{equation}
The Ward identity for Abelian theory takes the form:
\begin{equation}\label{240}
\frac{1}{\alpha} \partial^{y}_{\nu_{1}} \left[\frac{1}{i}
\frac{\delta {\rm Z}}{\delta J_{\nu_{1}}(y)}\right] = - \int
d^{4}z J_{\nu_{1}}(z) \partial_{\nu_{1}}^{z}D(y-z) \cdot {\rm Z}.
\end{equation}
Applying the operator $\partial_{\nu_{2}}^{q} \left[ \frac{1}{i}
\frac{\delta}{\delta J_{\nu_{2}}(q)} \right]$ to this equation,
and then the operator $\frac{1}{i^{2}} \frac{\delta}{\delta
J_{\mu}(x)}\frac{\delta}{\delta J_{\mu}(0)}$, we obtain:
\begin{eqnarray}
\frac{1}{\alpha} \partial_{\nu_{1}}^{y} \partial_{\nu_{2}}^{q}
\left[ \frac{1}{i^{2}} \frac{\delta^{2} {\rm Z}}{\delta
J_{\nu_{1}}(y) \delta J_{\nu_{2}}(q)} \right] = -\frac{1}{i}
\delta(y-q) \cdot {\rm Z} - \frac{1}{i} \int d^{4}z J_{\nu_{1}}(z)
\partial_{\nu_{1}}^{z} D(y-z) \partial_{\nu_{2}}^{q} \frac{\delta
{\rm Z}}{\delta
J_{\nu_{2}}(q)};\\
\frac{1}{\alpha} \partial_{\nu_{1}}^{y} \partial_{\nu_{2}}^{q}
\left[ \frac{1}{i^{4}} \frac{\delta^{4} {\rm Z}}{\delta
J_{\nu_{1}}(y) \delta J_{\nu_{2}}(q) \delta J_{\mu}(x) \delta
J_{\mu}(0)}\right]_{J=0} = \frac{1}{i^{3}} \delta(y-q)
\frac{\delta^{2}{\rm Z}}{\delta J_{\mu}(x) \delta
J_{\mu}(0)}|_{J=0} -
\\ \nonumber - \frac{1}{i^{3}} \times \left\{
\partial_{\mu}^{x} D(y-x) \partial_{\nu_{2}}^{q} \frac{\delta^{2} {\rm Z}}{\delta J_{\nu_{2}}(q) \delta J_{\mu}(0)} -
\partial_{\mu}^{y} D(y) \partial_{\nu_{2}}^{q} \frac{\delta^{2}{\rm Z}}{\delta J_{\nu_{2}}(q) \delta
J_{\mu}(x)}\right\}_{J=0}
\end{eqnarray}
Setting $q=y$ and integrating over $y$, we get:
\begin{eqnarray}
{\rm Q'} = - \frac{i}{2 \alpha} \frac{1}{i} \left( -(2\pi)^{4}
(\delta(0))^{2} {\rm Q \; +  N} \int d^{4}y
\partial_{\mu}^{y} D(y) \partial_{\nu_{2}}^{y} <T A_{\nu_{2}}(y)A_{\mu}(x)> -\right.\\ \nonumber \left.-\; {\rm N} \int d^{4}y \partial_{\mu}^{x} D(y-x) \partial_{\nu_{2}}^{y} <T A_{\nu_{2}}(y) A_{\mu}(0)>
\right).
\end{eqnarray}
Here, of course, by \;$\delta(0)$\; we should understand the
regularization $\delta (0) = (2\pi)^{-4} \cdot\Omega$ ($\Omega$ is
the volume of momentum space) and on having carried out the
calculations, we take the limit $\; \Omega \rightarrow \infty$.
Integrating by parts in the last two integrals and using the Ward
identity for the two-point Green's function
$$\partial_{\mu}^{x}\partial_{\nu}^{y} <T(A^{\mu}(x)A^{\nu}(y))> =
- \alpha \delta(x-y)$$ we obtain the result:
\begin{equation}\label{250}
{\rm Q'} = -\frac{{\rm N}}{2\alpha} \left(-(2\pi)^{4}
(\delta(0))^{2} <T(A_{\mu}(x)A^{\mu}(0))> - 2\alpha D(x)\right)
\end{equation}
Analogously we get
\begin{equation}\label{260}
{\rm N'} =  \frac{1}{2 \alpha} (2\pi)^{4} (\delta(0))^{2} {\rm N}
\end{equation}
Combining the last two equations, we have:
\begin{equation}\label{270}
{\rm Q' \cdot N - N' \cdot Q} = D(x) {\rm N^{2}},
\end{equation}
and due to (\ref{220}):
\begin{equation}\label{280}
\frac{d {\rm C}}{d\alpha} = D(x).
\end{equation}
As $C(x)$ is the contraction of the identity (\ref{90}) with
respect to the Lorentz indices, it is clear, that $G(p^{2})$
doesn't depend on $\alpha$.\newline

\textbf{Appendix 2.}

\emph{Impossibility of building a Wilson OPE in noncommutative
field theory.}

Let us demonstrate the impossibility of building a Wilson OPE in
noncommutative quantum field theory. We consider the simplest
example - the expansion for $T \phi(x)\phi(0)$ as $x \rightarrow
0$ in the case of $\phi \star \phi \star \phi \star \phi$-theory.
It is clear, that in the noncommutative field theory in principle
there could be two different expansions: those with commutative or
noncommutative condensates, i.e.:

\emph{Variant 1}
\begin{equation}\label{102}
T \phi (x) \phi (0) = C_{0}(x) \cdot 1 + C_{1}(x) \cdot [\phi(0)
\cdot \phi(0)] + ...
\end{equation}

\emph{Variant 2}
\begin{equation}\label{103}
T \phi (x) \phi (0) = C_{0}(x) \cdot 1 + C_{1}(x) \cdot [\phi(0)
\star \phi(0)] + ...
\end{equation}

Neither of these variants is realized. Let's consider a matrix
element over one-particle states $<p|T \phi(x) \phi(0)|k>$ in the
lowest order\footnote{Feynman rules for noncommutative
$\phi^{4}$-theory can be found, for instance, in ~\cite{arefeva}}
($\sim g$):
\begin{eqnarray}\label{100}
\int d^{4}x e^{iqx} <p|T \phi(x) \phi(0)|k> \sim \frac{g}{3}
\frac{1}{q^{2}+m^{2}} \frac{1}{(\omega-q)^{2}+m^{2}} \times \\
\nonumber \times \left(\cos{[(\omega-q)\times q]} \cos{[p\times
k]} + \cos{[(\omega-q)\times p]} \cos{[q\times k]} + \right.\\
\nonumber \left.+ \cos{[(\omega-q)\times k]} \cos{[q\times
p]}\right),
\end{eqnarray}
where $\omega = k-p$. As $q \rightarrow \infty$ we get the
following asymptotic behavior:
\begin{eqnarray}\label{104}
\underset{q \rightarrow \infty}{Asymp} \int d^{4}x e^{iqx} <p|T
\phi(x) \phi(0)|k> \sim \frac{g}{3} \frac{1}{q^{4}}
 \cdot
(\cos{[(\omega-q)\times q]} \cos{[p\times k]} + \\ \nonumber +
\cos{[(\omega-q)\times p]} \cos{[q\times k]} +
\cos{[(\omega-q)\times k]} \cos{[q\times p]}) \equiv A(q,k,p,
\omega)
\end{eqnarray}
As it is easy to see, this function $A(q,k,p,\omega)$ cannot be
decomposed into a product of two functions, depending on the large
momentum $q$ and small external momenta $p$ and $k$, i.e.
\begin{equation}\label{101}
A(q,k,p,\omega)  \neq B(q) \cdot C(k,p),
\end{equation}
whereas such representation would be necessary for a Wilson
expansion, both (\ref{102}) or (\ref{103}) (which is clear
directly from the form of these expansions).

For comparison let's consider the commutative limit $\xi
\rightarrow 0$ of the amplitude (\ref{100}) or the asymptotic
formula (\ref{104}):
\begin{equation}\label{105}
\underset{\xi \rightarrow 0}{\lim} A(q,k,p,\omega) \sim
\frac{g}{q^{4}},
\end{equation}
which is factorized in the form (\ref{101}) if we set:
\begin{eqnarray}
B(q) = \frac{g}{q^{4}}; C(k,p)=1.
\end{eqnarray}
Similar ideas were presented in the work ~\cite{zamora}.
\newline \newline This work has been partially supported by RFBR grants,
the grant for the support of leading scientific schools and the program "Theoretical problems of mathematics".

\end{document}